\documentclass[prl,twocolumn,english,nofootinbib,preprintnumbers]{revtex4}
\usepackage{hyperref}
\usepackage[T1]{fontenc}
\usepackage{amsmath}
\usepackage{amssymb}
\usepackage{babel}
\usepackage[latin1]{inputenc}
\usepackage{latexsym}

\usepackage{ifpdf}
\ifpdf 
  \DeclareGraphicsExtensions{.pdf,.png}
  \usepackage{pgf}
\else 
  \usepackage{graphicx}
  \DeclareGraphicsExtensions{.eps,.png}
  \usepackage{pgf}
  \usepackage{pstricks}
\fi

\usepackage{epstopdf}
\graphicspath{{./}}

\newcommand{\dd}{\textrm{d}}  

\begin{document}
\preprint{HD-THEP-09-05}
\title{Viable Singularity-Free f(R) Gravity Without a Cosmological Constant}
\author{Vinícius Miranda\thanks{viniciusvmb@if.ufrj.br}, Sergio E. Jor\'as\thanks{joras@if.ufrj.br}, Ioav Waga\thanks{ioav@if.ufrj.br}}
\affiliation{Instituto de F\'{\i}sica, Universidade Fe\-de\-ral do
Rio de Janeiro, Caixa Postal 68528, Rio de Janeiro, RJ 21941-972,
Brazil}
\author{Miguel Quartin\thanks{quartin@mporzio.astro.it}}
\affiliation{INAF/Osservatorio Astronomico di Roma, V. Frascati 33,
00040 Monteporzio Catone, Roma, Italy \\
Universit\`a  di Milano-Bicocca, Dip. Fisica ``G. Occhialini'', P.le Scienze 3, 20126 Milano, Italy \\
Institut für Theoretische Physik, Universität Heidelberg, Philosophenweg 16, 69120 Heidelberg, Germany}
\date{\today}

\begin{abstract}
Several authors have argued that self-consistent $f(R)$ gravity
models distinct from $\Lambda $CDM are almost ruled out. Confronting
such claims, we present a particular two-parameter $f(R)$ model
that: (a) is cosmologically viable and distinguishable from $\Lambda
$CDM; (b) is compatible with the existence of relativistic stars;
(c) is free of singularities of the Ricci scalar during the
cosmological evolution and (d) allows the addition of high curvature
corrections that could be relevant for inflation.
\end{abstract}

\maketitle

\emph{Introduction}. Since the discovery of cosmic acceleration, more than a decade ago, considerable effort has been devoted in cosmology to understand what is the physical mechanism responsible for it. A relic cosmological constant $\Lambda $, even though arguably the simplest explanation and in good accordance with observations, faces some theoretical difficulties (mainly due to the cosmic coincidence problem and related fine-tuning~\cite{Quartin08}) that have motivated an intense search for alternatives. These can be divided into two main conceptual approaches, both involving the introduction of new degrees of freedom (see for instance~\cite{Uzan}): either one modifies the left hand side of Einstein's equations (modified gravity) or one adds a new term to the energy momentum tensor, arguably associated with a new fundamental field not directly related to gravity.

Special attention to the former approach has been given in the last
five years. In particular, $f(R)$ gravity theory, 
due to its simplicity, received the main focus (for a recent review,
see~\cite{Faraoni} and references therein). This approach amounts to
writing the action as \vspace{-0.0 cm}
\begin{equation}
    S=\int {d^{4}x\sqrt{-g}}\left[ \frac{1}{16\pi G}f(R) + \mathcal{L}_{mat}
    \right] ,  \label{acao}
    \vspace{-0.0 cm}
\end{equation}
where $f(R)=R+\Delta (R)$, $R$ is the Ricci scalar and $\Delta (R)$
is an arbitrary function. General Relativity (GR) without a
cosmological constant is obtained in the special case in which
$\Delta (R)$ is identically zero. Although a great deal of effort
has been employed to develop this approach, it appeared to be a
difficult challenge to build a new Lagrangian that does not spoil
the successes of GR --- one that passes solar system tests,
describes the early universe, allows a matter-dominated phase
followed by an accelerating attractor~\cite{Amendola} --- and, at
the same time, do not suffer from curvature
singularities~\cite{Frolov}. The presence of singularities may have
devastating consequences and could forbid, for instance, the
formation of relativistic stelar objects such as neutron
stars~\cite{Maeda}.


\emph{Singularity-Free f(R) Model}. Several popular $f(R)$ models investigated in the literature are generalized by the following expression \vspace{-0.0 cm}
\begin{equation}
    f(R)=R-R_S \beta \left\{ 1-\left[ 1+\left( \frac{R}{R_{\ast}}\right) ^{n}
    \right] ^{-\frac{1}{\beta }}\right\} .  \label{geral}
    \vspace{-0.0 cm}
\end{equation}
For instance, choosing $\beta =-1$ we obtain the models presented in~\cite{Rn}; for $\beta =1$ we recover the model proposed in~\cite{HuSa}; for $n=2$ we get the $f(R)$ function discussed in~\cite{Staro}. In this letter we consider the special case in which $n=1$ and we take the limit $\beta \rightarrow \infty $. In this limit~\eqref{geral} can be recast as (rewriting $R_S$ as
$\alpha R_{\ast}$) \vspace{-0.0 cm}
\begin{equation}
    f(R)=R-\alpha R_{\ast}\ln \left( 1+\frac{R}{R_{\ast}}\right) ,  \label{f}
    \vspace{-0.0 cm}
\end{equation}
where $\alpha $ and $R_{\ast}$ are free positive parameters. Notice that the above function satisfies the stability conditions~\cite{PoSi}: (a) $f_{RR}:= \text{d}^{2}f/\text{d} R^{2}>0$ (no tachyons~\cite{Faraoni06}); (b) $f_{R}:= \text{d} f / \text{d} R>0$ (no ghosts) for $\alpha <(\tilde{R}/R_{\ast}+1)$, where $\tilde{R} $ is the value of the Ricci scalar at the final accelerated fixed point; and (c) $\lim_{R\rightarrow \infty }\Delta /R=0$ and $\lim_{R\rightarrow \infty } \Delta _{R}=0$ (GR is recovered at early times). Above and henceforth, $\Delta _{R} := \text{d}\Delta /\text{d} R$.



Starting from the action~\eqref{acao}, one obtains the equation of motion for $f(R)$: \vspace{-0.0 cm}
\begin{equation}
    f_{R}R_{\mu \nu }-\nabla _{\mu }\nabla _{\nu }f_{R}+\left( \Box f_{R}-\frac{1%
    }{2}f\right) g_{\mu \nu }=8\pi GT_{\mu \nu },  \label{eqf}
    \vspace{-0.0 cm}
\end{equation}
the trace of which is given by
\begin{equation}
    \Box f_{R}=\frac{8\pi G}{3}T+\frac{1}{3}\left( 2f-f_{R}R\right) ,
    \label{trace}
    \vspace{-0.0 cm}
\end{equation}
where $T$ is the trace of the energy-momentum tensor. We now introduce the scalar degree of freedom (d.o.f.) $\chi :=f_{R}$ and write the equations based on the mapping from $f(R)$ gravity (with positive first and second derivatives) onto Brans-Dicke scalar-tensor theory with parameter $\omega =0$. The resulting field equation is
\begin{equation}
    \Box \chi =\frac{\text{d}V}{\text{d}\chi }-\mathcal{F}\,,  \label{chieq2}
    \vspace{-0.0 cm}
\end{equation}
with the force term given by $\mathcal{F}:=-(8\pi G/3)\,T$ and
\vspace{-0.1cm}
\begin{equation}
    \frac{\text{d}V\!\left( R(\chi )\right) }{\text{d}\chi }:=\frac{1}{3}\left(
    2f-f_{R}R\right) .  \label{dVdphi}
    \vspace{-0.0 cm}
\end{equation}

\begin{figure}[t]
    \includegraphics[width=8.0cm,height=5.2cm]{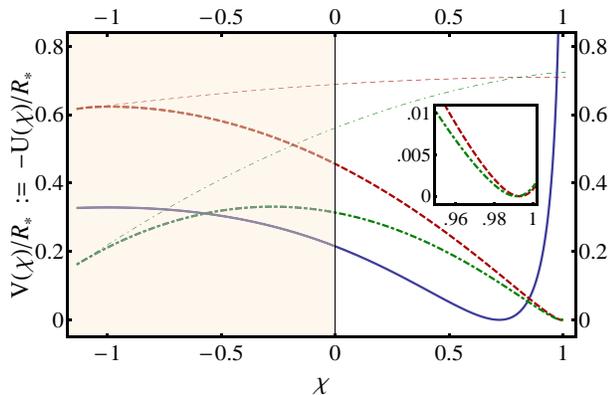}
    \caption{$V(\chi)/R_{\ast}:= - U(\chi)/R_{\ast}$ for different models: ours, with $\alpha=2$ (blue solid line),
    Starobinski's for \{n=2, $\lambda =2$\} [see~\cite{Staro}] (red, dashed line) and Hu \& Sawicki's for \{n=2, $m^2=1$,
    $c_1/c_2 = 2$\} [see~\cite{HuSa}] (green dot-dashed line). The physically interesting region is $0<\chi<1$.
    For the multi-valued potentials only the lower lines are physical.}
    \label{fig:V-of-X}
\end{figure}

When applying the model~\eqref{f} to a spatially homogeneous and isotropic universe the scalar d.o.f. $\!\!\!\!$ becomes
\begin{align}
    \chi [R(t)]=1-\frac{\alpha R_{\ast}}{R(t)+R_{\ast}}.  \label{scalarfrolov}
\end{align}
and the d'Alembertian in~\eqref{chieq2} is effectively just a
time-derivative: $\Box \equiv - \partial^2 / \partial t^2 - 3H\partial / \partial t$; our
choice for the metric signature is $(-,+,+,+)$. It is
straightforward to see that $\,\chi \rightarrow 1^{-}\,$ as
$\,R\rightarrow \infty $, which points out the same singularity
\cite{Frolov} featured in previous models~\cite{HuSa,Staro,Appleby}.
Inverting the relation~(\ref{scalarfrolov}) and
integrating~(\ref{dVdphi}) we find that (up to a constant)
\vspace{-0.1cm}
\begin{equation}
    \frac{3V(\chi )}{R_{\ast}}=-\alpha (2\chi -3)\ln \left( \frac{\alpha }{1-\chi }
    \right) +\left( \chi -1\right) \left( \frac{\chi -3}{2}-\alpha \right) .
    \label{vchi}
    \vspace{-0.0 cm}
\end{equation}
Note that since~\eqref{scalarfrolov} defines a one-to-one relation
between $\chi $ and $R$, the potential $V(\chi )$ is well-defined
and not multi-valued, contrary to the models
in~\cite{HuSa,Rn,Staro}. Fig.~\ref{fig:V-of-X} depicts the
potential for $\alpha = 2$, as well as typical potentials derived
from models~\cite{HuSa,Staro}. Taking the limit $\chi \rightarrow
1^{-}$ we find that
\begin{align}
    V(\chi \rightarrow 1^{-})\approx \frac{\alpha R_{\ast}}{3}\ln \left( \frac{\alpha }{1-\chi }\right) \,\rightarrow \,+\infty
    \vspace{-0.0 cm}
\end{align}
which shows the presence of an infinite barrier at $\chi =1$ \linebreak[4] that prevents the singularity discussed in \cite{Frolov} to be
reached.

We can understand this result in a more intuitive way by making use of the well-known duality between $f(R)$ and scalar-tensor theories: a conformal transformation of the metric can cast the Lagrangian from the Jordan into the Einstein frame, converting the scalar d.o.f.~$\chi$ into a canonical scalar field~$\tilde{\chi}:=-\sqrt{3/16\pi G} \ln{\chi}$~\cite{brax}. The field equation for $\tilde{\chi}$ has the same structure of~\eqref{chieq2}, but with the following potential
\vspace{-0.1cm}
\begin{equation}
    V_E\left( R\big(\tilde{\chi}\big)\right) =\frac{1}{16\pi G}\frac{R\Delta _{R}-\Delta }{(1+\Delta _{R})^{2}}.  \label{frolov2}
    \vspace{-0.0 cm}
\end{equation}
All the discussion above, regarding the presence of an infinite
barrier, applies to $V_E$ as well. Note that since $1+\Delta_{R}>0$ (stability condition (b)), the numerator of Eq.~(\ref{frolov2}) is the only factor that can
make the potential diverge as $R\rightarrow \infty $.
In~\cite{abhadev} the singularity was avoided by introducing an
extra high-curvature term $\alpha R^{n\,(>1)}$ in the model
investigated in~\cite{Staro}. It is easy to see why that kind of
correction works: in that case, the numerator in~(\ref{frolov2}) is
itself proportional to $R^{n}$. Nevertheless, such term cannot be
used, at the same time, both to avoid the singularities and to
generate inflation~\cite{Maeda2}. This is not the case of the model
investigated in this letter, since it is not necessary to include
such terms to avoid the two singularity problems, as we have shown
above (for the case discussed in \cite{Frolov}) and will show bellow
for the case discussed in \cite{Maeda}.

Notice that two different singularity-free classes of $f(R)$ are
possible: we can pick a function $\Delta $ such that either
$\lim_{R\rightarrow \infty }R\Delta _{R}=\infty $ or
$\lim_{R\rightarrow \infty }\Delta =-\infty $ holds. In the former
case, $\Delta $ can even become constant as $R\rightarrow \infty $
--- which actually happens in the models previously
mentioned~\cite{HuSa,Staro} --- but it should do so slowly, thus
keeping the divergence of $R\Delta _{R}$, which does not happen on
those models. The model~(\ref{f}) belongs to the latter case.
Another interesting example of this class is \vspace{-0.0 cm}
\begin{equation}
    f(R)=R-\alpha R_{\ast}\left( 1+\frac{R}{R_{\ast}}\right) ^{n}
    \vspace{-0.0 cm}
\end{equation}
with $\alpha >0$ , $R_{\ast}>0$ and $n\in (0,1)$. Although
preliminary tests indicate that this model is cosmologically viable,
it carries an explicit positive cosmological constant, in direct
contrast to~\eqref{f}.

We further remark that the potential~(\ref{frolov2}) derived
from~\eqref{f} generates a Yukawa-like force which is fully
compatible with the Chameleon mechanism~\cite{brax,KhoWel}. In other
words, the mass of the $\chi$ field is large (small) when the
background matter density is large (small). This mass dependence on
the local environment explains how this extra (or fifth) force can
have cosmological implications while at the same time evading
detection by local gravity experiments.


\emph{Relativistic Stars}. The authors of~\cite{Maeda} argue that
the very existence of relativistic stars poses a strong constraint
on $f(R)$ gravity theories. For the models studied in that paper, it
was not possible to evolve the metric from inside a star up to large
spatial scales and match the de Sitter solution asymptotically. We
show below that this divergence is circumvented by model~\eqref{f}
and, therefore, does not represent a general feature of $f(R)$
models. For the sake of clarity, we follow the classical-mechanics
analogy used in \cite{Maeda} and the necessary definitions. We
consider a static and spherically symmetric metric and write the
d'Alembertian in~\eqref{chieq2} as $\Box \equiv
\partial^2 /
\partial r^2  +  (2/r) \partial /
\partial r $ in spherical coordinates; we are assuming a Minkowski background for a moment.
In this case, Eq.~\eqref{chieq2} can be seen as the equation of
motion of a classical particle of unit mass (albeit one whose
``time'' coordinate is our spatial coordinate $r$) submitted to both
an external and frictional forces. Therefore \vspace{-0.0 cm}
\begin{equation}
    \frac{\dd^{2}\chi }{\dd r^{2}}+\frac{2}{r}\frac{\dd\chi }{\dd r}=\tilde{\mathcal{F}}+\mathcal{F}_{U},  \label{class}
    \vspace{-0.0 cm}
\end{equation}
where $\tilde{\mathcal{F}}:=-\mathcal{F}$ and $\mathcal{F}_{U}:=-\dd
U / \dd\chi$ are, respectively, the ``force'' due to the trace of
the energy-momentum tensor (non vanishing inside the star) and the
``force'' due to the potential $U(\chi )=-V(\chi )$, see
Eq.~\eqref{vchi}  and Fig.~\ref{fig:V-of-X}. Again, the change in
sign is just a consequence of the fact that now it is the spatial
(instead of time) dependence of $\chi$ which is the most relevant.

For the models analyzed in~\cite{Maeda}, there was no solution which
would describe a particle going uphill pulled by the force
$\tilde{\mathcal{F}}$ (while still inside the star) and stop at the top of
the potential at $r\rightarrow \infty $, which would correspond to
the de Sitter metric. The particle would either return and reach the
singularity at $\chi =1$ (where $R\rightarrow \infty $) or overshoot
the potential towards $\chi =0$ (which would also lead to a
singularity, for instance, in the Kretschmann scalar $K:=R^{\alpha
\beta \mu \nu }R_{\alpha \beta \mu \nu }$). Fairly enough, $U(\chi
)$ diverges at $\chi =1$, as in all other models~\cite{HuSa,Staro}.
As we will show below, the advantage here is a well-behaved solution
fully compatible with relativistic stars embedded in a de Sitter
universe.

\begin{figure}[t]
\includegraphics[width=5.0cm,height=8.2cm,angle=-90]{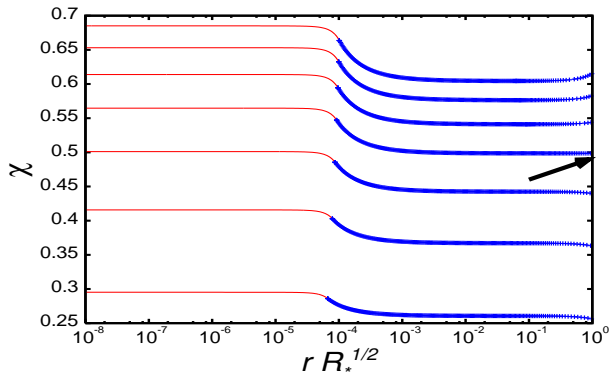}
\caption{The field $\protect\chi$ with $\protect\alpha=1.2$,
$p_c=0.3 \protect\rho_0$ and $R_c$ varying from $1\times
10^{-8}\protect\rho_0 $ to $4\times 10^{-8}\protect\rho_0$. The
arrow points out the solution that stops at the maximum of the
potential at $r\rightarrow \infty$.  The (red) thin lines indicate
the region inside the star.} \label{chi}
\end{figure}
Let us now determine the full evolution of the $\chi$ field. As
previously mentioned, we start from a static and spherically
symmetric metric \vspace{-0.0 cm}
\begin{equation}
    ds^{2}=-N(r)\,dt^{2}+\frac{1}{B(r)}\,dr^{2}+r^{2}\,d\Omega ^{2}
    \label{metric}
    \vspace{-0.0 cm}
\end{equation}
and assume a constant energy-density star whose energy-momentum
tensor is given by \linebreak[4]$\;T_{\mu }^{~\nu
}=\mathrm{diag}(-\rho _{0},p(r),p(r),p(r))$. The initial conditions
at $\;r_{i}=10^{-8}R_{\ast}^{-1/2}$, i.e, close to the center of the
star, are given by $\;N(r_{i})=1+N_{2}\,r_{i}^{2}$,
$\;B(r_{i})=1+B_{2}\,r_{i}^{2}\;$,
$\;p(r_{i})=p_{c}+p_{2}\,r_{i}^{2}/2\;$ and $\;\chi (r_{i})=\chi
_{c}\left( 1+C_{2}\,r_{i}^{2}/2\right)$. The coefficients $\,N_{2}$,
$\,B_{2}$, $\,p_{2}\,$ and $\,C_{2}\,$ can be written in terms of
$\;\rho_{0}=2\times 10^{8}\Lambda _{\mathrm{eff}}\;$ and of the
central values $\;p_{c}=0.3\rho _{0}$, $\;R_{c}=10^{-8}\rho _{0}$,
$\;V(\chi _{c})$ and $\;\dd V/\dd\chi (\chi _{c})$. The effective
value of the cosmological constant is given by $\;\Lambda
_{\mathrm{eff}}=R_{1}/4$, where $\,R_{1}\,$ is the value of the
Ricci scalar when $\;\dd V/\dd\chi =0$. We refer the reader to the
original paper \cite{Maeda} for the full set of equations. Energy
conservation provides an important relation between $p(r)$ and
$N(r)$ inside the star. We evolve the system $\;\{p,B,\chi ,\dd\chi
/\dd r\}\;$ from $\,r_{i}\,$ up to the radius $\,\mathcal{R}\,$ of
the star (defined by $\,p(\mathcal{R})=0$) where we require
continuity of the variables. From then on we evolve the system
$\;\{N,B,\chi ,\dd\chi /\dd r\}\;$ until $\,r=R_{\ast}^{-1/2}$
(cosmological scales).

We show in Figure~\ref{chi} the
behavior of the field $\chi $ for different values of initial
conditions. Note that some trajectories do not get past the top of
the potential and return towards the singularity at $\chi =1$ (top 3
curves) while others (3 lowest ones) overshoot and go towards $\chi
=0 $ and one (indicated by an arrow) stops right at the maximum. We
recall that this solution was obtained without any high-curvature
correction. It is obviously an issue of fine tuning the initial
conditions to stop exactly there. Another remarkable feature of this
model is the absence of singularity in $K$ as $\chi $ decreases
below the peak of its potential.


\emph{A Promising Model}. A viable cosmological model must start
with a radiation-dominated universe and have a saddle point
matter-dominated phase followed by an accelerated epoch as a final
attractor. We can formally state such criteria if we use the
parameters \linebreak[4] $m:=Rf_{,RR}/f_{R}\;$ and $\;\mathfrak{r}
:=-Rf_{R}/f\,$. We refer the reader to the original
paper~\cite{Amendola} for a full discussion on this subject. An
early matter-dominated epoch of the universe can be achieved if
$m(\mathfrak{r}\approx -1)\approx 0^+$ and $\dd
m/\dd\mathfrak{r}(\mathfrak{r}\approx -1)>-1$. Furthermore, a
necessary condition for a given model to reach a late-time
accelerated phase is $0<m(\mathfrak{r}\approx -2)\leq 1$. The model
(\ref{f}) satisfies both constraints for $\alpha >1$ regardless of
$R_{\ast}$.

\begin{figure}[t]
\includegraphics[width=8.2cm,height=5.2cm]{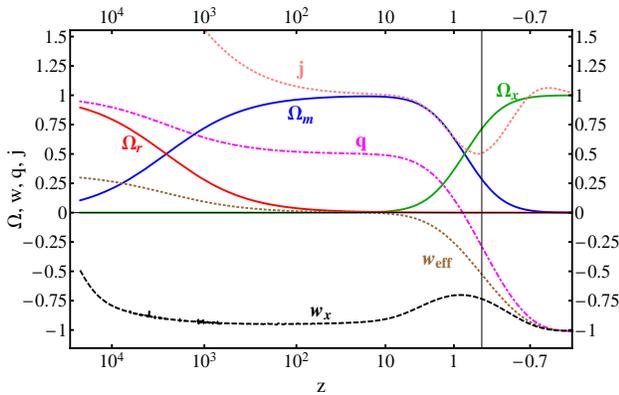}
\caption{Cosmological evolution of the densities $\Omega _{m}$, $\Omega _{r}$%
, $\Omega_{c}$ (solid lines), the deceleration factor $q$
(dot-dashed line),
the jerk $j$ (dotted line) , the equation of state parameters $w_{x}$ and $%
w_{\mathrm{eff}}$ (dashed and dotted lines, respectively), for $\protect%
\alpha = 2$. } \label{cosmo}
\end{figure}

Using~(\ref{eqf}), we obtain the modified Einstein's equations below
for a homogeneous universe filled with matter energy density $\rho
_{m}$ (baryons and cold dark matter) and radiation energy density
$\rho _{r}$: \vspace{-0.0 cm}
\begin{align}
    3H^{2} =& \;8\pi G\left( \rho _{m}+\rho _{r}\right) \,+\,\left(f_{R}R-f\right)/2 \,-\,3H\dot{f}_{R}\,+ \notag \\
    &\quad +\,3H^{2}(1-f_{R}) \\
    -2\dot{H} =& \; 8\pi G\left(\rho _{m}+4\rho _{r}/3\right) + \ddot{f}_{R} - H\dot{f}_{R}-2\dot{H}(1-f_{R}),
\end{align}\vspace{-0.4cm}

\noindent where a dot corresponds to derivative with respect to $t$,
$H\equiv \overset{\cdot }{a}/a$ and $a(t)$ is the scale factor. From
the equations above, we can define $\rho _{x}$, $p_{x}$ and
$w_{x}:=p_{x}/\rho _{x}$, respectively the energy density, pressure
and the equation-of-state parameter of the so-called ``curvature
fluid'': \vspace{-0.0 cm}
\begin{align}
    8\pi G\rho _{x} :=&\;\left( f_{R}R-f\right)/2 \,-\, 3H\dot{f}_{R} \,+\, 3H^{2}(1-f_{R}) \label{rhox}\\
    8\pi Gp_{x} := & \;\ddot{f}_{R} \,+\, 2H\dot{f}_{R} \,-\, (2\dot{H}+3H^{2})(1-f_{R})\,+ \notag \\
    &\qquad +\,(f-f_{R}R)/2\, \label{px}.
\end{align}
\vspace{-0.6cm}

\noindent These definitions are such as to guarantee that the
curvature fluid is conserved and only minimally coupled to matter
and radiation~\cite{Perrota}. We also define the relative densities
$\Omega _{i}$ (where $i$ stands for either radiation, matter or
curvature) $\Omega _{i}:=8\pi G\rho _{i}/3H^{2}$.

In Figure~\ref{cosmo} we plot the behavior of $\Omega _{m}$, $\Omega _{r}$, $%
\Omega _{x}$, the deceleration parameter
$q:=-\ddot{a}a/{\dot{a}}^{2}$, the jerk
$j:=\overset{\ldots}{a}a^{2}/{\dot{a}}^{2}$, the equation of state
parameters for the curvature fluid $w_{x}$ and for the effective
fluid $w_{\mathrm{eff}}:=p_{\mathrm{tot}}/\rho _{\mathrm{tot}}\equiv
(p_{r}+p_{x})/(\rho _{m}+\rho _{r}+\rho _{x})$, all of which can be
written in terms of known variables $R$, $H^{2}$ and $\rho _{i}$. In
Fig.~\ref{cosmo} we can clearly distinguish the radiation-dominated
era when $q\simeq 1$ (and $j\simeq 3$, not shown), followed by a
transient domination by matter ($q\simeq 1/2$ and $j \simeq 1$), the
current accelerated expansion ($q<0$) and the final de Sitter
attractor ($q=-j=-1$). We find similar results for different initial
conditions and parameters, indicating what seems to be an absence of
fine tuning. We remark that the $w_{x}$ curve in Fig.~\ref{cosmo} is
noisy in the early universe since at that time $\rho _{x}$ is too
small and the numerical calculation of $w_{x}$ becomes inaccurate.

We point out that there is some residual arbitrariness in defining
$\rho_x$ and $p_x$ even if one is only interested in conserved and
minimally coupled fluids. The one we follow, together with the
definition of $\Omega_i$, is convenient for comparison with GR-based
interpretations of observations. As another consequence of Eqs.
(\ref{rhox}) and (\ref{px}), $w_x$ neither crosses $-1$ nor diverges
at low redshift in contrast with~\cite{amendola2}, where slightly
different definitions are adopted. Note, however, that  observable
quantities like $H$ and $\rho_m$ are well-defined and in fact, using
either definition, have the same cosmological evolution.

\emph{Conclusions}. We have shown that some recent results in the
literature regarding divergences in $f(R)$ theories are not as
general as previously thought. In fact, even a compact,
two-parameter lagrangian like the one in~\eqref{f} can evade the
aforementioned singularities. Observational constraints on this
model are under investigation and the results will be published
elsewhere.

We thank Luca Amendola, Philippe Brax and Ribamar R. R. Reis for
helpful discussions in the course of this work. VM and IW
acknowledge financial support from the Brazilian Research Agency
CNPq. MQ thanks OAR, U. Milano-Bicocca and ITP U. Heidelberg for support.
V. M.  would like to use this opportunity to honor his old friend the late Professor Marcos Azevedo da Silveira.

\end{document}